\documentclass{article}

\usepackage{arxiv}

\usepackage[utf8]{inputenc} % allow utf-8 input
\usepackage[T1]{fontenc}    % use 8-bit T1 fonts
\usepackage[hidelinks]{hyperref}       % hyperlinks
\usepackage{url}            % simple URL typesetting
\usepackage{booktabs}       % professional-quality tables
\usepackage{amsfonts}       % blackboard math symbols
\usepackage{nicefrac}       % compact symbols for 1/2, etc.
\usepackage{microtype}      % microtypography
\usepackage{lipsum}		% Can be removed after putting your text content
\usepackage{graphicx}
\usepackage[numbers]{natbib}
\usepackage{doi}
\usepackage{bm}
\usepackage{bbm}

\usepackage{setspace}
\title{A prediction interval for the population-wise error rate}

\author{{Remi Luschei} \\
	Competence Center for Clinical Trials Bremen \\
	Institute for Statistics\\
	University of Bremen\\
	\texttt{rluschei@uni-bremen.de} \\
	\And
	{Werner Brannath} \\
	Competence Center for Clinical Trials Bremen\\
	Institute for Statistics \\
	University of Bremen\\
	\texttt{brannath@uni-bremen.de} \\
}

\renewcommand{\headeright}{}
\renewcommand{\undertitle}{}
\hypersetup{
pdftitle={A prediction interval for the population-wise error rate},
pdfauthor={Remi Luschei, Werner Brannath},
}

\usepackage{amsmath, amsfonts, mathtools}
\usepackage{url}
\usepackage[titletoc,toc,page]{appendix}
\usepackage{multirow}
\newcommand{\Pop}{\mathcal{P}}
\newcommand{\R}{\mathbb{R}}

\renewcommand{\P}{\mathbb{P}}
\newcommand{\PWER}{\text{PWER}}
\newcommand{\FWER}{\text{FWER}}

\newcommand{\bzero}{\bm{0}}
\newcommand{\bc}{\bm{c}}
\newcommand{\bpi}{\bm{\pi}}
\newcommand{\bSigma}{\bm{\Sigma}}

\begin{document}
\maketitle 

\begin{abstract}
We construct an asymptotic prediction interval for the population-wise error rate (PWER), which is a multiple type I error criterion for clinical trials with overlapping patient populations. The PWER is the probability that a randomly selected patient will receive an ineffective treatment. It must usually be estimated due to unknown population strata sizes, such that only an estimate can be controlled at the given significance level. We apply the delta method to find a prediction interval for the resulting true PWER, we demonstrate by simulations that the interval has the required coverage probability, and illustrate the approach with real data examples.
\end{abstract}


\section{Introduction} \label{sec1}

\citet{brannath} have introduced the population-wise error rate (PWER) as a new type I error measure for studies that investigate treatments in multiple, possibly overlapping populations. Let $\Pop_1, \dots, \Pop_m, m \geq 2$, denote these populations. Consider the null hypotheses $H_i: \theta_i \leq 0$, $i \in I = \{1, \dots, m\}$, where $\theta_i = \theta(\Pop_i, T_i)$ denotes the effect of an investigational treatment $T_i$ in comparison to a control treatment $C$ in population $\Pop_i$. We define the disjoint population strata $\Pop_J = \left( \cap_{j \in J} \Pop_j \right) \backslash \left( \cup_{k \in I \setminus J} \Pop_k \right)$ for all $J \subseteq I$, that include all patients affected by the treatments indexed in $J$. For each $\Pop_J$, let $\pi_J$ denote its relative prevalence among the total patient population. The population-wise error rate is defined by \[\PWER = \sum_{J \subseteq I} \pi_J \P(\text{reject any true $H_j$ for $j \in J$}).\] We assume that the hypotheses $H_i$ are tested with test statistics $Z_i$ that have a known joint distribution, with corresponding critical values denoted by $c_i \in \R$. Then the PWER can be controlled (and exhausted) at a given significance level $\alpha \in (0,1)$ by selecting these critical values such that 
\begin{align} \label{eq1}
\PWER = \sum_{J \subseteq I} \pi_J \P\left(\cup_{j \in J \cap I_0} \{Z_j > c_j\} \right) = \alpha,
\end{align}
where $I_0 = \{i \in I: \theta_i \leq 0\}$ is the index set of the true null hypotheses.
As we have described in \cite{luschei1}, the prevalences $\pi_J$ are often unknown in practice. But approximate PWER-control may then still be possible by plugging in their maximum-likelihood-estimator. Consider a sample of $N$ patients, with $n_J$ patients belonging to stratum $\Pop_J$ for each $J \subseteq I$. We can model the strata sample sizes $(n_J)_{J \subseteq I}$ by a multinomial distribution with parameters $N$ and $\boldsymbol{\pi} = (\pi_J)_{J \subseteq I}$, where the corresponding maximum-likelihood estimator of $\boldsymbol{\pi}$ is
\begin{align} \label{eq2}
\hat{\boldsymbol{\pi}} = (\hat{\pi}_J)_{J \subseteq I}, \quad \text{with} \quad \hat{\pi}_J = \frac{n_J}{N}.
\end{align}
From Equation (\ref{eq1}) we can thus obtain a vector of estimated critical values $\hat{\bc} = \bc(\hat{\bpi}) = (\hat{c}_1, \dots, \hat{c}_m)$ such that the estimated PWER equals $\alpha$. We have shown in \cite{luschei1} that the resulting true PWER (which uses the true prevalences, and the components of $\hat{\bc}$ as critical values) then converges almost surely to $\alpha$ for $N \to \infty$. Moreover, we showed by simulation studies that the true PWER closely approximates $\alpha$ in finite samples and is almost perfectly controlled for different realistic choices of $N$, such as $N=100, 250, 500$. However, we may want or see the need to quantify the small variations in a single study. To this end, we will derive in this paper asymptotic prediction intervals for the true PWER for individual studies.

\section{Computation of the PWER} \label{sec:pwer}

In all strata $\Pop_J$ and for all treatments $T$ given in $\Pop_J$, we assume normally distributed observations with possibly heterogeneous means $\mu_{J, T}$ and variances $\sigma_{J,T}^2$. The null hypotheses $H_i$ can then often be tested via Gaussian- or $t$-tests, as described in \cite{luschei1}. When the variances $\sigma_{J,T}^2$ are assumed to be known, the maximal PWER is computed by \[\PWER(\bc) = \sum_{J \subseteq I} \pi_J (1- \Phi_{\bzero, \bSigma_J}(\bc_J)), \quad \text{for} \quad  \bc_J = (c_j)_{j \in J},\] where $\Phi_{\bzero, \bSigma_J}$ denotes the cumulative distribution function of the multivariate normal distribution with mean $\bzero = (0, \dots, 0)^T \in \R^{|J|}$ and the correlation matrix $\bSigma_J = (\Sigma_{ij})_{i,j \in J}$ given by
\begin{align} \label{corr}
\Sigma_{ij} =  \frac{1}{\sqrt{ V_iV_j}}&\sum_{J \subseteq I:\, i,j \in J} \frac{n_{J,T_i}\sigma^2_{J, T_i}}{n_{i, T_i}n_{j,T_i}} \mathbbm{1}(T_i = T_j) + \frac{n_{J,C} \sigma^2_{J, C}}{n_{i,C} n_{j,C}}, \quad i\neq j, \\
&\text{where} \quad V_i = \sum_{J \subseteq I: \, i \in J} \frac{n_{J, T_i}\sigma^2_{J, T_i}}{n_{i,T_i}^2}  + \frac{n_{J,C}\sigma^2_{J,C}}{n_{i,C}^2}. \notag
\end{align}
Here $n_{J, T}$ and $n_{i, T}$ denote the number of patients receiving treatment $T \in \{T_i, C\}$ in $\Pop_J$ and $\Pop_i$, respectively, and $\mathbbm{1}(T_i = T_j)$ is the indicator function that equals 1 if $T_i = T_j$ and 0 otherwise.

In case that the variances $\sigma_{J,T}^2$ are unknown but homogeneous ($\sigma_{J,T}^2=\sigma^2$ for all $J$ and $T$), the distribution function $\Phi_{\bzero, \bSigma_J}$ can be replaced by the distribution function of the multivariate $t$-distribution, with $\bSigma_J$ as the scale matrix, and with $N-s$ degrees of freedom, where $s$ denotes the number of combinations of $J$ and $T$. If the variances $\sigma_{J,T}^2$ are unknown and heterogeneous, then the joint distribution of the test statistics is not available in closed form. This remains an unsolved problem \cite{hasler}. The distribution may be approximated from simulated data, as we proposed in \cite{luschei1} in Section~4.7.

However, the above normal and $t$-distributions are only valid under the assumption of homogeneous null effects, i.e.\ when the strata-wise effects $\theta_{J, T_i}$, for all $J\subseteq I$, go in the same (positive or negative) direction. If strata effects may have opposite signs, i.e.\ a so-called qualitative effect heterogeneity \cite{wang, gabler} occurs, the joint distribution of the test statistics is unknown under the null hypotheses. We have adressed this problem in \cite{luschei2} and proposed a bootstrap-based approach to approximate the distribution.

\section{Derivation of the prediction interval}

We derive an asymptotic prediction interval for the true PWER, that is obtained by plugging the estimated critical values $\hat{\bc} = \boldsymbol{c}(\hat{\bpi})$ into the function $\PWER(\bc)$ (see Section~\ref{sec1}). We utilize the fact that the maximum-likelihood estimate of the multinomial distribution %from Equation (\ref{eq2}) 
is asymptotically normal, i.e., \[\sqrt{N}(\hat{\bpi}-\bpi) \xrightarrow{d} N(\bzero, \mathbf{R}),\] where $\mathbf{R} = \text{diag}(\bpi) - \bpi \bpi^T$ \cite{akoto}. By the delta method, the expression
\begin{align*}
\sqrt{N}\left(\text{true PWER}-\alpha\right) = \sqrt{N}\left(\PWER(\hat{\bc})-\PWER(\bc)\right)
\end{align*}
converges in distribution to the univariate normal distribution $\mathrm{N}(0, \gamma^2)$ with $\gamma^2 = \nabla_{\bpi} \PWER(\bc)^T \mathbf{R} \nabla_{\bpi} \PWER(\bc)$, where $\nabla_{\bpi} \PWER(\bc)$ denotes the gradient of $\PWER(\bc)$ with respect to $\bpi$. It takes the form $\nabla_{\bpi} \PWER(\bc)=\left(F_J(\bc_J) -1\right)_{J \subseteq I}$, where $F_J$ denotes the joint distribution function of the test statistics $(Z_j)_{j \in J}$. The calculation of the gradient is detailed in Appendix~\ref{app1}. Consequently, an asymptotic ($1-\alpha'$) prediction interval for the true PWER is given by 
\begin{align} \label{pi}
\left[\alpha - \frac{z_{\alpha'/2}\gamma}{\sqrt{N}}, \alpha + \frac{z_{\alpha'/2}\gamma}{\sqrt{N}} \right],
\end{align}
where $\alpha$ is the chosen significance level and $z_{\alpha'/2}$ denotes the $(1-\alpha'/2)$ quantile of the standard normal distribution. We note that the value of $\gamma$ will be unknown in practice due to the unknown true prevalences, but it may be estimated via plug-in of $\hat{\bpi}$.

As we have seen in Section~\ref{sec:pwer}, an exact analytical expression of the gradient of the PWER is not available in the cases with unknown, heterogeneous strata-treatment-wise variances, or when the assumption of homogeneous null effects in the strata does not hold. Therefore, in these cases the gradient needs to be simulated using Monte Carlo or bootstrap methods, as described in \cite{luschei1, luschei2}. We will apply these in Section~\ref{sec:sim-unknown}.

\subsection{Accounting for a minimal prevalence} \label{sec:min}

In \cite{luschei1} we propose to introduce a minimal strata prevalence $\pi_\text{min}$ at the estimation of the prevalence vector $\bpi = (\pi_J)_{J \subseteq I}$ to account for small strata whose existence is known, but in which no patients could be recruited due to their small size. The idea is to include all strata-specific family-wise error rates $\FWER_J = 1- F_J(\mathbf{c}_J)$ in the PWER with a weight of at least $\pi_\text{min}$ in order to limit future risks for patients also in the small strata. In this case, we would be interested in finding a prediction interval for the true PWER, where the prevalences have been transformed in the same way as in the prevalence estimation.

Let $\mathcal{J}_\text{min} = \{J \subseteq I: \pi_J < \pi_\text{min}\}$. One possible transformation $\boldsymbol{\tilde{\pi}}=(\tilde{\pi}_J)_{J \subseteq I}$ would be
\begin{align} \label{trans1}
\tilde{\pi}_J = \pi_\text{min}, \quad \text{when } \pi_J < \pi_\text{min} \quad \text{and} \quad \tilde{\pi}_J = p \cdot \pi_J \text{ with }  p\coloneqq \frac{1-\sum_{J\in \mathcal{J}_\text{min}} \pi_\text{min}}{1-\sum_{J \in \mathcal{J}_\text{min}}\pi_J} \quad \text{otherwise},
\end{align}
that is, all prevalences that are smaller than the specified minimum $\pi_\text{min}$ are increased to $\pi_\text{min}$, and all others are scaled down proportionally to their original size. The gradient of the PWER then takes the form
\[\frac{d}{d\pi_J} \PWER(\bc)= 0, \quad \text{when } \pi_J < \pi_\text{min} \quad \text{and} \quad \frac{d}{d\pi_J} \PWER(\bc) = p \cdot (F_J(\bc_J) -1), \quad \text{when } \pi_J > \pi_\text{min}.\]
Note that the PWER is not differentiable for $\pi_J = \pi_\text{min}$ under this transformation. However, this is a quite rare and unlikely scenario in practice.

Nevertheless, to fix this issue one could set 
\begin{align} \label{trans2}
\tilde{\pi}_J = \frac{\pi_J + \pi_\text{min}}{1 + n_S \pi_\text{min}}, \quad \text{where $n_S$ denotes the total number of strata}.
\end{align}
The gradient of the PWER is then $\nabla_{\bpi} \PWER(\bc)=\left((F_J(\bc_J) -1)/(1 + n_S \pi_\text{min})\right)_{J \subseteq I}$. We will investigate and compare the two different transformations in Section~\ref{sec:sim-min}.

\section{Simulation of the coverage probability}

We investigated the coverage probability of the estimated prediction interval in simulation studies across different configurations. All simulations were conducted in R and  can be reproduced with the script files available under \url{https://github.com/rluschei/pi-pwer}.

We considered different total sample sizes, $N \in \{250, 500, 1000\}$, which are realistic values for multi-population studies, and different numbers of populations ranging from $m = 2, \dots, 5$. The significance level for PWER control was chosen as $\alpha = 0.025$ as the tests are one-sided. The general procedure in all simulations is the following: We first set up some true strata prevalences $\bpi$ (e.g.\ equal prevalences for all strata) and repeatedly generate a vector of strata-wise sample sizes $\boldsymbol{n} = (n_J)_{J \subseteq I}$ from the multinomial distribution with parameters $N$ and $\bpi$. From $\boldsymbol{n}$ and $N$ we compute the prevalence estimator $\hat{\bpi}$ and plug it into the prediction interval, which is computed using Formula~(\ref{pi}). By counting the number of occurences where the true PWER lies in the resulting interval, divided by the number of simulation runs, we obtain the estimated coverage probability.

\subsection{Cases with known distribution} \label{sec:sim-known}

We first consider the different settings with known or unknown and possibly heterogeneous strata-treatment-wise variances, where the joint distribution of the test statistics is known (see Section~\ref{sec:pwer}):
\begin{itemize}
\item[A:] known + homogeneous variances
\item[B:] known + heterogeneous variances
\item[C:] unknown + homogeneous variances
\end{itemize}

We assume equal strata prevalences and that the treatments tested in the different populations are pairwise different ($T_i \neq T_j$ for $i\neq j)$ -- which influences on the correlation matrix (\ref{corr}) of the test statistics. In setting B, the known residual variances are randomly generated in each simulation run from independent uniform distributions between 0 and 1. In the settings A and C, this is not necessary since they cancel out in Formula~(\ref{corr}) in the homogeneous case.

Table~\ref{tab1} shows the resulting probability values for a targeted coverage probability of 0.95. We did a total of 10,000 simulation runs in every case. It turns out that the desired coverage is achieved with high accuracy in all configurations, as the observed values are consistently within a 0.003 range around the nominal level.

In addition, Table~\ref{tab1} reports the mean length of the simulated prediction intervals ($\text{length} \times 10^3$, in brackets). As expected, smaller values of $N$ lead to larger prediction intervals than higher values of $N$. For $m=4$ populations and $N=250$ patients, the mean length is $2.46 \cdot 10^{-3}$, which means that on average, true PWER values up to 0.02623 can plausibly be contained within the interval. For an increasing number of populations, no particularly large increase in interval length is observed. The complete distribution of the interval lengths across different study settings will be examined below. 

\begin{table}[ht]
\centering
\begin{tabular}{cc|ccccccccc}
\toprule
\multirow{2}{*}{$N$} &
\multirow{2}{*}{Setting} & 
\multicolumn{4}{c}{Coverage ($\text{length} \times 10^3$) of prediction interval} \\
 & &$m=2$ & $m=3$ & $m=4$ & $m=5$ \\
\midrule
250 & A & 0.9483 \,(2.10) & 0.9446 \,(2.42) & 0.9484 \,(2.46) & 0.9493 \,(2.39)\\
250 & B & 0.9474 \,(2.09)& 0.9481 \,(2.42)& 0.9495 \,(2.46)& 0.9507 \,(2.39)\\
250 & C & 0.9483 \,(2.10)& 0.9491 \,(2.42)& 0.9502 \,(2.46)& 0.9500 \,(2.38)\\
500 & A & 0.9476 \,(1.49)& 0.9518 \,(1.72)& 0.9499 \,(1.74)& 0.9484 \,(1.69)\\
500 & B & 0.9522 \,(1.48)& 0.9485 \,(1.71)& 0.9466 \,(1.74)& 0.9512 \,(1.69)\\
500 & C & 0.9476 \,(1.48)& 0.9488 \,(1.71)& 0.9512 \,(1.74)& 0.9492 \,(1.69)\\
1000 & A & 0.9517 \,(1.05)& 0.9497 \,(1.21)& 0.9513 \,(1.23)& 0.9488 \,(1.20)\\
1000 & B & 0.9516 \,(1.05)& 0.9495 \,(1.21)& 0.9531 \,(1.23)& 0.9500 \,(1.20)\\
1000 & C & 0.9517 \,(1.05)& 0.9516 \,(1.21)& 0.9512 \,(1.23)& 0.9536 \,(1.20)\\
\bottomrule
\end{tabular}
\vspace{10pt}
\caption{Simulated coverage probability and mean prediction interval $\text{length} \times 10^3$ in settings with equal strata prevalences and pairwise different treatments. $N$ = sample size, $m$ = number of populations}
\label{tab1}
\end{table}

In further simulations, we also considered the case of a single investigational treatment tested in all populations and obtained very similar results. We additionally investigated the precision of the prediction intervals for other choices of true prevalences. For example, we considered the setting already regarded in \cite{luschei1}, in which one stratum has a rather large prevalence of 0.5 while the others have equal prevalences, and again found results that were very similar to the equal prevalences case. All of these additional simulation results are provided in the accompanying Github repository. 

To give a representative summary of these investigations, Table~\ref{tab2} reports the distribution of further simulated coverage probabilities over 100 studies with varying prevalences. In each study, we defined the target populations $\Pop_i$ as the subgroup of patients where a certain binary biomarker is expressed. The probability of each biomarker being expressed in the overall patient population was drawn independently from a uniform distribution. All strata in which at least one biomarker was expressed were included, and their prevalences were computed as the product of the corresponding biomarker probabilities, and then normalized to sum to one. The number of simulation runs for each choice of prevalences was reduced to 1,000, so that the coverage probability could be computed for all of the 100 studies within a reasonable runtime. As shown in Table~\ref{tab2}, the resulting mean and median coverage probabilities are rather close to 0.95. However, in certain studies the coverage falls noticeably below the nominal level. This occurs particularly for lower numbers of populations and is usually associated with situations where individual strata have very small prevalences.

\begin{figure}[htbp]
  \centering
  \includegraphics[width=0.8\textwidth]{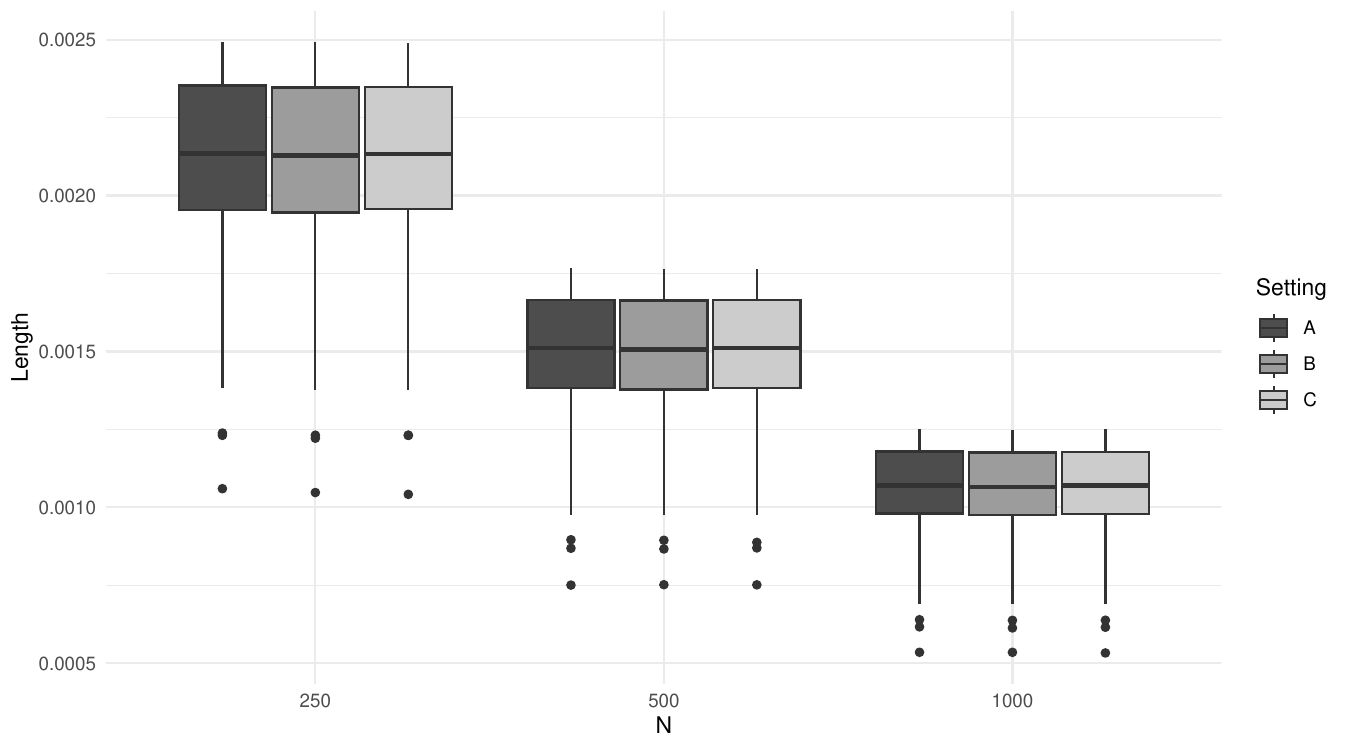}
  \caption{Distribution of the mean prediction interval lengths for 100 studies with $m=3$ populations, randomly generated prevalences and different population-wise treatments. $N$ = sample size.}
  \label{fig2}
\end{figure}

Table~\ref{tab2} also contains the average lengths of the prediction intervals, which are obtained as the means of the study specific lengths. We see that in the average, even smaller values are achieved than in the scenario of Table~\ref{tab1} with equal prevalences. To provide an overview of the variability of the lengths across the different studies, see Figure~\ref{fig2}, in which the entire distribution is plotted exemplarily for $m=3$ populations and different sample sizes. 

\begin{table}[ht]
\centering
\begin{tabular}{cc|cccccccccc}
\toprule
$m$ &
Setting & 
Mean &SD &Min &Q1& Med &Q3& Max & $\text{Length} \times 10^3$\\
\midrule
2 & A & 0.8690 & 0.1238 & 0.062 & 0.8555 & 0.9115 & 0.9380 & 0.960 & 1.20\\
2 & B & 0.8675 & 0.1261 & 0.070 & 0.8558 & 0.9070 & 0.9372 & 0.959 & 1.19\\
2 & C & 0.8690 & 0.1238 & 0.062 & 0.8555 & 0.9115 & 0.9380 & 0.960 & 1.19\\
3 & A & 0.9231 & 0.0395 & 0.731 & 0.9138 & 0.9360 & 0.9470 & 0.963 & 1.48 \\
3 & B & 0.9232 & 0.0397 & 0.736 & 0.9140 & 0.9380 & 0.9470 & 0.961 & 1.48\\
3 & C & 0.9229 & 0.0390 & 0.753 & 0.9132 & 0.9385 & 0.9472 & 0.960 & 1.48\\
4 & A & 0.9384 & 0.0210 & 0.835 & 0.9358 & 0.9450 & 0.9510 & 0.961 & 1.52 \\
4 & B & 0.9378 & 0.0211 & 0.847 & 0.9323 & 0.9450 & 0.9492 & 0.966 & 1.51 \\
4 & C & 0.9372 & 0.0216 & 0.855 & 0.9350 & 0.9430 & 0.9500 & 0.960 & 1.52 \\
5 & A & 0.9426 & 0.0123 & 0.892 & 0.9400 & 0.9450 & 0.9500 & 0.962 & 1.47 \\
5 & B & 0.9432 & 0.0104 & 0.906 & 0.9388 & 0.9435 & 0.9500 & 0.964 & 1.47 \\
5 & C & 0.9434 & 0.0122 & 0.888 & 0.9398 & 0.9450 & 0.9510 & 0.968 & 1.47 \\
\bottomrule
\end{tabular}

\vspace{10pt}
\caption{Distribution of the simulated coverage probability over 100 studies with randomly generated prevalences, $N=500$ patients, and different population-wise treatments. The last column gives the average length of the prediction intervals over all studies.}
\label{tab2}
\end{table}

To systematically examine the effect of small strata prevalences on the coverage probability, we considered a scenario in which the prevalence of the stratum $\Pop_I$ is set to $1/(2^{m+4}-16)$, i.e.\ one sixteenth of the proportion that a stratum would have if all strata were equal in size. For $m=2$, this corresponds to a prevalence of $0.0208$. The remaining strata are assumed to have equal prevalences. By definition, $\Pop_I$ is the intersection of all target populations $\Pop_i$ for $i \in I$ and, if it exists, corresponds to the stratum with the highest strata-wise error probability. We considered settings with $m=2$ populations, in which one individual stratum may exert a relatively strong influence. The results confirm the observation made before: For $N=250$ patients, the coverage probability is approximately 88 percent. As the sample size increases, however, it rises again, for example to almost 94 percent for $N=500$. Notably, the prediction intervals observed in these cases were shorter than those reported in the previously discussed scenarios (e.g.\ $8.18 \cdot 10^{-4}$ for $N=250$ and $5.90 \cdot 10^{-4}$ for $N=500$). Note that in cases with such small prevalences, a minimal strata prevalence might be applied (see Section~\ref{sec:min}). The simulation results for this situation are presented in Section~\ref{sec:sim-min}.

\subsection{Cases with unknown distribution} \label{sec:sim-unknown}

In this section, we consider the following scenarios, where the distribution of the test statistics is unknown and must be approximated (see Section~\ref{sec:pwer}):

\begin{itemize}
\item[D:] unknown + heterogeneous variances (but homogeneous null effects)
\item[E:] qualitative null effect heterogeneity (but unknown + homogeneous variances)
\end{itemize}

In these cases, in each simulation run, the gradient of the PWER and the true PWER (which both depend on the strata-wise sample sizes $\boldsymbol{n}$) must be simulated within another simulation subloop, as described in the two next paragraphs. Due to the increased computational effort needed to run those simulations, we therefore restrict our analysis to settings with $m=2$ populations and $N=250$ patients. 

In Case D, we apply and compare two different methods to estimate the critical values $\boldsymbol{\hat{c}}$. The first method uses a multivariate $t$-approximation based on degrees of freedom according to \citet{satterthwaite}, that has been described in detail in \cite{luschei1}, Section~4.7. In the second method, the critical values are computed using the following parametric bootstrap-method: In each iteration, a random sample is generated in every stratum under the global null hypothesis, i.e.\ from independent normal distributions with mean 0 and variance equal to the observed variance divided by the sample size of the stratum. The test statistics are computed from this sample, and for every given $\boldsymbol{c}$, the strata-wise error probabilities $\FWER_J=1-F_J(\boldsymbol{c}_J)$ are approximated by the empirical rejection frequencies in the stratum $\Pop_J$. To the thereby defined empirical PWER-function, a numerical root-finding method is applied to solve the equation $\PWER(\boldsymbol{\hat{c}}) = \alpha$ for $\boldsymbol{\hat{c}}$.

For both methods, the gradient of the PWER, as well as the true PWER, are also approximated from the resampling strategy described above. We obtain the gradient by taking the negative values of the $\FWER_J$, for all $J \subseteq I$, and the true PWER is obtained by taking the weighted mean of the $\FWER_J$, using the true prevalences as weights instead of the estimated prevalences. The number of resamples was set to $10^4$, respectively, for each of the $10^4$ generated sample size vectors $\boldsymbol{n}$. As before, we considered different settings with equal prevalences, one high and one low prevalence, as well as equal or pairwise different experimental treatments. The results can be found in Table~\ref{tab4}. We observe that, across all configurations, the Satterthwaite method achieves only low coverage probabilities, which are far away from the targeted coverage probability of 0.95. Under the bootstrap approximation, however, the targeted coverage is mostly attained with high accuracy, except in the already discussed cases with one very small prevalence (see Section~\ref{sec:sim-known}).

In Case E, each simulation run begins by setting up some heterogeneous treatment effects across the strata. Specifically, we randomly draw an effect for the stratum $\Pop_{\{1,2\}}$ (the intersection of the populations $\Pop_1$ and $\Pop_2$) from the uniform distribution on $(-1,1)$. The effects in the remaining strata are then computed so that, when weighted by the corresponding prevalences, the overall average treatment effects are zero in both $\Pop_1$ and $\Pop_2$ (this is a uniquely solvable system of linear equations). Next, we generate strata-specific sample means by drawing from normal distributions centered at the true means, with variance $\sigma^2$ divided by the stratum sample size. For simplicity, we set $\sigma = 0.5$ in all strata. A corresponding pooled sample variance is generated using a $\chi^2$-distribution. The critical values, the gradient and the PWER are then all obtained using the same bootstrap procedure: In each bootstrap iteration, the strata sample sizes are redrawn from a multinomial distribution with probabilities equal to the observed strata proportions. The test statistics are then resampled under the global null hypothesis, which is implemented by projecting the observed effects onto the linear space defined by the null hypotheses. Further details of the procedure can be found in \cite{luschei2}, Section~3.1. Finally, the strata-wise error probabilities are estimated in the same way as in Case D. A total of $10^4$ bootstrap repetitions were performed in all of the $10^4$ simulated study settings. The results of the simulations are shown in Table~\ref{tab4}. As can be seen, the nominal coverage probability of 0.95 is achieved with high accuracy in all settings for which this was already the case in the previously studied scenarios.

Finally, we note that the average observed lengths of the simulated prediction intervals are quite similar across the three methods and scenarios, and that they are of the same order of magnitude as in the cases with equal prevalences discussed in Section~\ref{sec:sim-known}.

\begin{table}[ht]
\centering
\begin{tabular}{cc|cc|c}
\toprule
\multirow{2}{*}{Prevalences} &
\multirow{2}{*}{Treatments} & 
\multicolumn{2}{c}{Setting D} &
\multicolumn{1}{c}{Setting E}\\
&& Satterthwaite & Bootstrap & Bootstrap\\
\midrule
equal & single &  0.4871 \,(1.90) & 0.9449 \,(1.89)& 0.9520 \,(1.93)\\
equal & different & 0.5328 \,(2.12)& 0.9443 \,(2.10)& 0.9510 \,(2.10)\\ 
one large & single &  0.4233 \,(1.65)& 0.9490 \,(1.64)& 0.9405 \,(1.81) \\
one large & different & 0.4830 \,(1.98)& 0.9496 \,(1.96)& 0.9359 \,(1.98)\\
one small & single &  0.2373 \,(0.94)& 0.7616 \,(0.93)& 0.8931 \,(1.09)\\
one small & different &  0.1436 \,(1.36)& 0.4866 \,(1.86)& 0.8943 \,(1.12)\\
\bottomrule
\end{tabular}
\vspace{10pt}
\caption{Simulated coverage probability and mean prediction interval $\text{length} \times 10^3$ in the settings D and E, where the distribution of the test statistics is unknown, with $m=2$ populations and $N=250$ patients. Prevalences: equal: equal in all strata, one large: 0.5 in $\Pop_I$ and equal in others, one small:  $1/(2^{m+4}-16)$ in $\Pop_I$ and equal in others. Treatments: single: $T_i = T$ for all $i\in I$, different: $T_i \neq T_j$ for all $i \neq j$}
\label{tab4}
\end{table}

\subsection{PWER with a minimal prevalence} \label{sec:sim-min}

In the settings with small prevalences described in Section~\ref{sec:sim-known}, we applied different minimal prevalence boundaries $\pi_\text{min}$ to the PWER, such as $1/(2^{m+1} -2)$ and $1/(2^{m+2}-4)$. These thresholds correspond, respectively, to the half size of a stratum and to one quarter of the size of a stratum, assuming all strata are equal in size. We implemented both transformations (\ref{trans1}) and (\ref{trans2}) presented in Section~\ref{sec:min}, and  investigated the coverage probability of the prediction interval for the transformed PWER. We
considered different settings with $N \in \{250, 500\}$ patients, $m \in \{2, \dots, 5\}$ populations and pairwise different treatments. The simulation results are included in Table~\ref{tab6}. It shows that the prevalence estimate (\ref{trans1}) applied to the PWER leads to higher coverage probabilities of the corresponding prediction interval than using no minimal prevalence ($\pi_\text{min} = 0$), while the coverage remains almost unchanged for the prevalence transformation (\ref{trans2}). But note that in both cases, introducing a minimal prevalence will reduce the maximal strata-wise error probability, which is the primary aim of introducing the minimal prevalence. This has been described in \cite{luschei1}, Section~4.6.

Furthermore, we observe that using a minimal prevalence may reduce the length of the corresponding prediction intervals, depending on the value of $\pi_\text{min}$ and on how strongly the true, small and unknown prevalences are elevated. In particular, higher minimal prevalences lead to shorter prediction intervals, even though the accuracy of the resulting prevalence estimates does not improve and in fact decreases further as $\pi_\text{min}$ increases. The concrete lenghts observed in the scenarios described here can be found in Table~\ref{tab10} in Appendix~\ref{app2}. Comparing the two transformations, the estimator based on transformation (\ref{trans2}) tends to give shorter prediction intervals than with transformation (\ref{trans1}), with the difference becoming more pronounced as $\pi_\text{min}$ increases.

In transformation (\ref{trans1}), the problem is that the PWER is not differentiable in cases where $\pi_\text{min}$ equals the true prevalence $\pi_J$ of a stratum. Although this corresponds to a very unlikely scenario in practice, we conducted simulations for scenarios where we intentionally set $\pi_\text{min} = \pi_J$ for some $J$, and tried both using $\frac{d}{d\pi_J} \PWER(\bc)= 0$, as well as $\frac{d}{d\pi_J} \PWER(\bc) = p \cdot (F_J(\bc_J) -1)$ at the gradient calculation (using the notation of Section~\ref{sec:min}). We observed no substantial differences in the overall coverage probabilities and interval lengths.

\begin{table}[ht]
\centering
\begin{tabular}{cc|cccc|cccc} 
\toprule 
\multirow{3}{*}{$N$} & 
\multirow{3}{*}{$\pi_\text{min}$} & 
\multicolumn{4}{c|}{Estimate (\ref{trans1})} & \multicolumn{4}{c}{Estimate (\ref{trans2})} \\ \cmidrule(lr){3-6} \cmidrule(lr){7-10} & & \multicolumn{4}{c|}{$m$} & \multicolumn{4}{c}{$m$} \\ 
	 & 			& 2 		 & 3 		  & 		 4 & 		  5 & 		2 & 		 3 & 			4 & 5 \\ \midrule 
250 & 0 		& 0.8798 & 0.9478 & 0.9451 & 0.9508 & 0.8798 & 0.9478 & 0.9451 & 0.9508 \\ 
250 & $1/(2^{m+2}-4)$ 	& 0.9945 & 0.9608 & 0.9582 & 0.9538 & 0.8798 & 0.9514 & 0.9450 & 0.9464 \\ 
250 & $1/(2^{m+1}-2)$ 	& 0.9945 & 0.9675 & 0.9595 & 0.9631 & 0.8798 & 0.9486 & 0.9493 & 0.9492 \\ 
500 & 0 
    & 0.9374 & 0.9490 & 0.9500 & 0.9474 
    & 0.9374 & 0.9490 & 0.9500 & 0.9474 \\

500 & $1/(2^{m+2}-4)$ 
    & 1.0000 & 0.9620 & 0.9531 & 0.9534 
    & 0.9374 & 0.9475 & 0.9506 & 0.9495 \\

500 & $1/(2^{m+1}-2)$ 
    & 1.0000 & 0.9719 & 0.9597 & 0.9554 
    & 0.9374 & 0.9491 & 0.9461 & 0.9499 \\
 \bottomrule
\end{tabular}
\vspace{10pt}
\caption{Simulated coverage probability for the PWER with a minimal prevalence $\pi_\text{min}$, using the prevalence estimates (\ref{trans1}) and (\ref{trans2}) from Section~\ref{sec:min}. The true prevalences are $1/(2^{m+4}-16)$ in $\Pop_I$ and equal in the other strata. $N$ = sample size, $m$ = number of populations}
\label{tab6}
\end{table}

\section{Real data application}

\begin{figure}[htbp]
  \centering
  \includegraphics[width=0.45\textwidth]{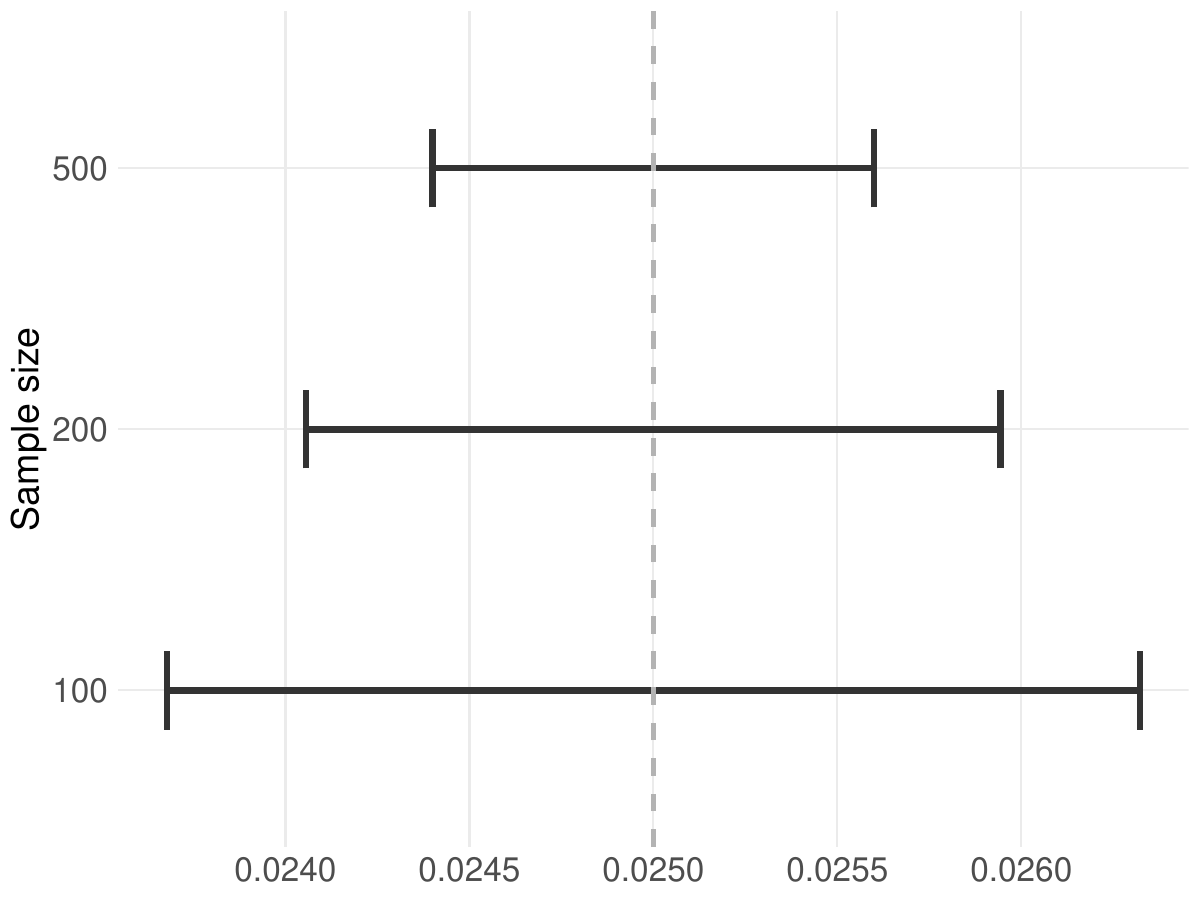}
  \hfill
  \includegraphics[width=0.45\textwidth]{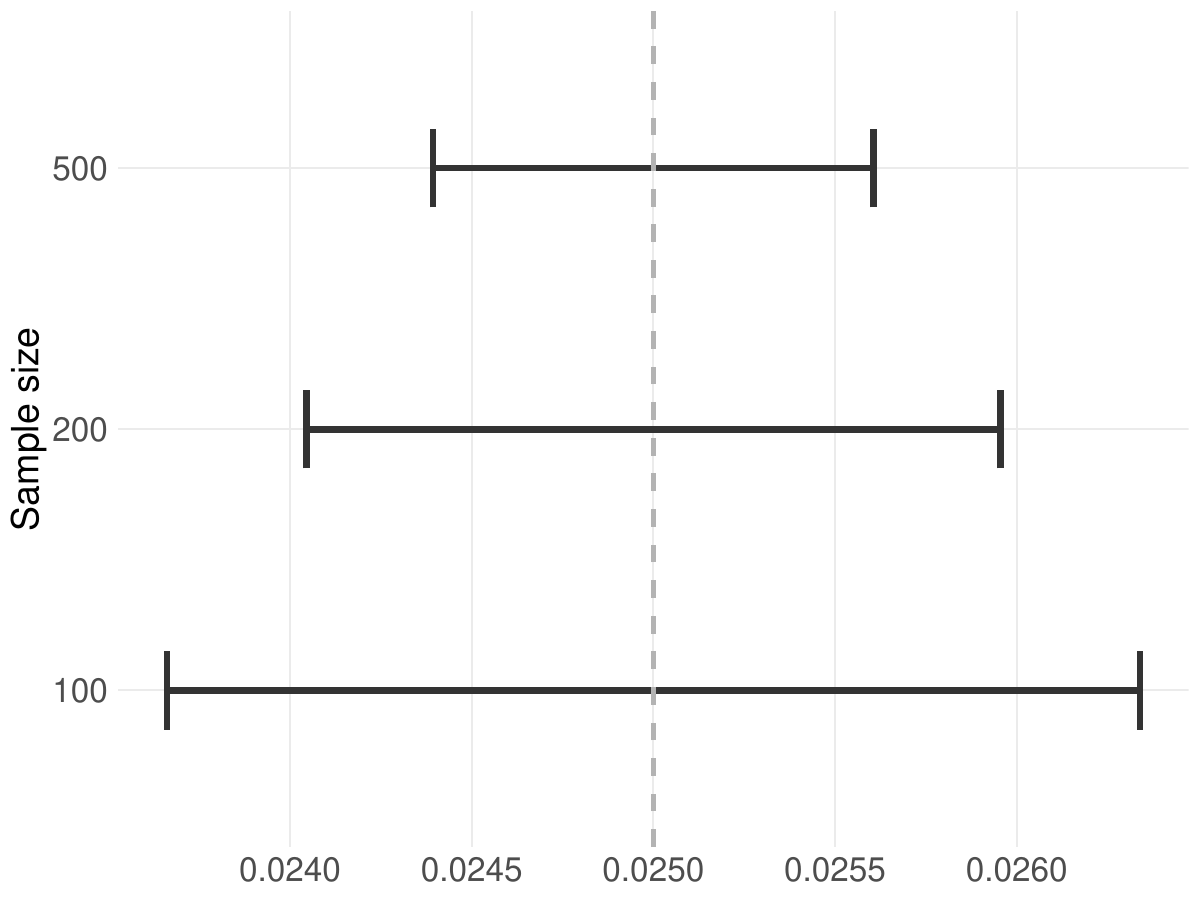}
  \caption{Asymptotic 95\% prediction intervals for the true PWER in the real data example from \citet{kesselmeier} for the random allocation strategy (left plot) and the pragmatic allocation strategy (right plot), and significance level $\alpha = 0.025$}
  \label{fig1}
\end{figure}

We illustrate the prediction interval for the PWER in the real data example used in \cite{luschei1}. The example is based on a data set created by \citet{kesselmeier}, using results of the MAXSEP study \cite{brunkhorst} that compared the effect of meropenem to the effect of a combination therapy with moxifloxacin and meropenem in patients with severe sepsis. \citet{kesselmeier} generated 1,000 bootstrap samples of sizes $N \in \{100, 200, 500\}$ to compare two patient allocation strategies in an umbrella trial design with two overlapping patient populations: (1) a pragmatic strategy, which assigns patients eligible for both subtrials to the currently smaller one, and (2) random allocation of these patients. For both strategies, in each resample we computed the two boundaries of the prediction interval for the PWER, and took their arithmetic mean over all resamples, respectively. The resulting mean prediction intervals, built from the mean lower and upper boundaries, are shown in Figure~\ref{fig1}. We can see that they do not differ significantly between the two strategies, and are getting narrower with increasing sample size, as expected.  

\section{Discussion}

Using the delta method, we derived asymptotic prediction intervals for the true value of the PWER that results when estimating the strata-wise sample sizes with the maximum-likelihood estimator of the multinomial distribution. Thus, in a practical situation, when the observed sample sizes are given, it is possible to provide a prediction of the achieved level of PWER control, which corresponds to the average risk for future patients of being exposed to inefficient treatments. In the simulations, we observed that the expected coverage probability is generally achieved with very high accuracy, with only few exceptions that have been discussed. It was furthermore shown that the prediction intervals are generally sufficiently narrow to allow for meaningful conclusions.

We call the resulting intervals \emph{predicition} intervals and not \emph{confidence} intervals since they provide a prognosis about the expected PWER-value in individual studies, where the strata-wise sample sizes have been realized. The PWER is here seen as a random variable that depends on the sample sizes which follow a multinomial distribution. The prediction intervals depend on the sample sizes only through the correlations of the test statistics, which enter through the standard deviation $\gamma$ in Formula~(\ref{pi}), but they also depend on the true prevalences $\boldsymbol{\pi}$. In practice, the true prevalences are in turn estimated using the realized sample sizes, which still gives reliable coverage results, as has been shown.

In Section~\ref{sec:sim-unknown}, it was shown that under the $t$-approximation with degrees of freedom according to the Satterthwaite formula, only low coverage probabilities are achieved, likely because this approximation is rather inaccurate for the chosen sample sizes. In \cite{luschei1}, we already observed that under this approximation the true PWER has a larger variance than in other settings in which the distribution of the PWER is known. In contrast, the bootstrap procedure proposed in Section~\ref{sec:sim-unknown} yielded more accurate coverage probabilities. We therefore compared the distribution of the true PWER under these two approximations and found that under the bootstrap, the true PWER is substantially closer to the nominal significance level. For example, in the setting from Section~4.7 in \cite{luschei1}, with $m=2$ populations and $N=500$ patients, the Satterthwaite approximation yielded a mean PWER of 0.02516 with a standard deviation of $1.3 \cdot 10^{-3}$, wheras the bootstrap procedure resulted in a mean PWER of $0.02501 \pm 3.8 \cdot 10^{-4}$ (with $10^4$ bootstrap repetitions). For these reasons, we would recommend using the bootstrap procedure for the determination of the critical values, when homogeneous null effects can be assumed and the residual variances are unknown and heterogeneous (Case D in Section~\ref{sec:sim-unknown}). 

If heterogeneous strata effects may occur under the null hypotheses, another boostrap procedure that explicitely accounts for this possibility should be employed (Case E in Section~\ref{sec:sim-unknown}). Otherwise, strong control of the PWER generally cannot be guaranteed. The reason is that assuming zeros as the strata-wise expected means under the null hypotheses, as is done in the homogeneous case (and in the bootstrap method mentioned above), does not necessarily represent a least favorable parameter configuration for the PWER in this situation. The same issue also applies to control of similar error rates such as the family-wise error rate. More details are provided in \cite{luschei2}, where also a corresponding bootstrap approach has been proposed.

\bibliographystyle{unsrtnat-max3}
\bibliography{references}

\appendix
\renewcommand{\thesubsection}{\Alph{subsection}}
\section*{Appendix} 
\addcontentsline{toc}{section}{Appendix}

\subsection{Calculation of the gradient of the true PWER} \label{app1}

By the multivariate chain rule, the gradient of $\PWER(\bc)$ is given by
\begin{align*}
\nabla_{\bpi} \PWER(\bc) 
    &= \frac{\partial \PWER}{\partial \bc}(\bpi, \bc) \, \frac{\partial \bc}{\partial \bpi}(\bpi),
\end{align*}
where $\frac{\partial \PWER}{\partial \bc}$ and $\frac{\partial \bc}{\partial \bpi}$ are the Jacobians of the $\PWER$ and $\bc$, respectively. The Jacobian of $\bc$ is obtained from the implicit function theorem by
\begin{align*}
\frac{\partial \bc}{\partial \bpi}(\bpi) 
    &= - \left( \frac{\partial \PWER}{\partial \bc}(\bpi, \bc) \right)^{-1}
       \left( \frac{\partial \PWER}{\partial \bpi}(\bpi, \bc) \right).
\end{align*} All in all, we obtain
\begin{align*}
\nabla_{\bpi} \PWER(\bc) 
    &= -\frac{\partial \PWER}{\partial \bpi}(\bpi, \bc) = -\frac{\partial}{\partial \bpi} \sum_{J \subseteq I} \pi_J (1- F_J(\bc_J)) = \left(F_J(\bc_J) -1\right)_{J \subseteq I}.
\end{align*}

\subsection{Further simulation results} \label{app2}

Table~\ref{tab10} reports the lengths of the prediction intervals with minimal prevalences (see Section~\ref{sec:sim-min}).

\begin{table}[ht]
\centering
\begin{tabular}{cc|cccc|cccc}
\toprule
\multirow{3}{*}{$N$} & \multirow{3}{*}{$\pi_\text{min}$}
& \multicolumn{4}{c|}{Estimate (\ref{trans1})}
& \multicolumn{4}{c}{Estimate (\ref{trans2})} \\
\cmidrule(lr){3-6} \cmidrule(lr){7-10}
& & \multicolumn{4}{c|}{$m$} & \multicolumn{4}{c}{$m$} \\
& & 2 & 3 & 4 & 5 & 2 & 3 & 4 & 5 \\
\midrule
250 & 0
& 0.82 & 2.06 & 2.28 & 2.29
& 0.82 & 2.06 & 2.28 & 2.29 \\

250 & $1/(2^{m+2}-4)$
& 0.77 & 2.01 & 2.25 & 2.28
& 0.62 & 1.61 & 1.80 & 1.82 \\

250 & $1/(2^{m+1}-2)$
& 0.72 & 1.94 & 2.22 & 2.26
& 0.50& 1.32& 1.49 & 1.51 \\

500 & 0
& 0.59 & 1.46 & 1.61& 1.62
& 0.59 & 1.46 & 1.61 & 1.62\\

500 & $1/(2^{m+2}-4)$
& 0.56 & 1.42 & 1.60 & 1.61
& 0.45 & 1.14 & 1.28 & 1.29 \\

500 & $1/(2^{m+1}-2)$
& 0.52 & 1.38 & 1.57& 1.60
& 0.36 & 0.93 & 1.06 & 1.07 \\
\bottomrule
\end{tabular}

\vspace{10pt}
\caption{Observed mean lengths $\times 10^{-3}$ of the prediction intervals with a minimal prevalence $\pi_\text{min}$, using the prevalence estimates (\ref{trans1}) and (\ref{trans2}) from Section~\ref{sec:min}. The true prevalences are $1/(2^{m+4}-16)$ in $\Pop_I$ and equal in the other strata. $N$ = sample size, $m$ = number of populations}
\label{tab10}
\end{table}

\end{document}